\documentclass [iop]{emulateapj}
\usepackage{natbib}
\begin{document}

\title{The Mass-Richness Relation of MaxBCG Clusters from Quasar Lensing Magnification using Variability}

\author{Anne H. Bauer\altaffilmark{1}, Charles Baltay\altaffilmark{2}, Nancy Ellman\altaffilmark{2}, Jonathan Jerke\altaffilmark{2}, David Rabinowitz\altaffilmark{2}, Richard Scalzo\altaffilmark{2,3}}
\email{bauer@ieec.uab.es}
\altaffiltext{1}{Institut de Ci\`encies de l'Espai, CSIC/IEEC, F. de Ci\`{e}ncies, Torre C5 par-2, Barcelona 08193, Spain}
\altaffiltext{2}{Yale University, Department of Physics, P.O. Box 208120, New Haven, CT 06520-8120, USA}
\altaffiltext{3}{Research School of Astronomy and Astrophysics, The Australian National University, Mount Stromlo Observatory, Cotter Road, Weston ACT 2611, Australia}

\begin{abstract}

Accurate measurement of galaxy cluster masses is an essential component not only in studies of cluster physics, but also for probes of cosmology.  However, different mass measurement techniques frequently yield discrepant results.  The SDSS MaxBCG catalog's mass-richness relation has previously been constrained using weak lensing shear, Sunyaev-Zeldovich (SZ), and X-ray measurements.  The mass normalization of the clusters as measured by weak lensing shear is $\gtrsim25\%$ higher than that measured using SZ and X-ray methods, a difference much larger than the stated measurement errors in the analyses.  
We constrain the mass-richness relation of the MaxBCG galaxy cluster catalog by measuring the gravitational lensing magnification of type I quasars in the background of the clusters.  
The magnification is determined using the quasars' variability and the correlation between quasars' variability amplitude and intrinsic luminosity.  
The mass-richness relation determined through magnification is in agreement with that measured using shear, confirming that the lensing strength of the clusters implies a high mass normalization, and that the discrepancy with other methods is not due to a shear-related systematic measurement error.  
We study the dependence of the measured mass normalization on the cluster halo orientation.  As expected, line-of-sight clusters yield a higher normalization;  however, this minority of haloes does not significantly bias the average mass-richness relation of the catalog.

\end{abstract}


\section{Introduction}

Galaxy clusters are powerful probes of cosmology, as they are the most massive collapsed structures in the universe.  The mass function of galaxy clusters is sensitive to the matter density $\Omega_{M}$, the amplitude of matter fluctuations $\sigma_{8}$, and the density and evolution of dark energy.  Large catalogs of homogeneously selected clusters can therefore be used to constrain these parameters \citep[e.g.,][]{rozo10, wen10, sehgal11}.  Individual clusters can also provide important cosmological information.   Hierarchical structure formation precludes the formation of very massive clusters at high redshift;  it is unclear if recent measurements of distant large clusters are compatible with hierarchical mass assembly in $\Lambda$CDM cosmology \citep[see, e.g.,][]{mortonson11, foley11, yaryura11, jee11}.  Accurate cluster mass measurements  are critical to each of these analyses.  

The mass of galaxy clusters can be estimated using several methods.  Weak lensing shear techniques involve measuring the total projected mass of a cluster by quantifying the gravitational distortion of the shapes of background galaxies (for reviews of weak lensing theory, see \citealt{bartelmann01} and \citealt{schneider04}.  For descriptions of some common weak lensing measurement techniques, see \citealt{ksb, refregier03, miller07}).  Thermal Sunyaev-Zeldovich (SZ; \citealt{sz}) analyses measure the change in the cosmic microwave background (CMB) due to inverse Compton scattering between electrons in the intra-cluster medium (ICM) and CMB photons.  The amplitude of the SZ signal is determined by the electron number density and temperature.  By assuming a pressure profile for the clusters, one can convert the SZ signal into a cluster mass measurement \citep[see][for a review]{birkinshaw99}.  X-ray observations also constrain the mass of a cluster, by measuring the thermal bremsstrahlung emission of electrons in the ICM;  given assumptions about the dynamical state of the cluster, the X-ray luminosity can be converted into a mass measurement \citep[e.g.,][]{pratt09}.  These techniques often struggle to agree when applied to the same clusters \citep[e.g.,][]{limousin10, huang11, morandi11}, although some systems do yield very good agreement \citep[e.g.,][]{zhang10, israel10, lerchster11}).

The MaxBCG catalog is an optically selected set of 13,823 galaxy clusters detected in the Sloan Digital Sky Survey \citep{maxbcg}.   The average mass profile of the MaxBCG clusters has been extensively studied using weak lensing shear \citep{sheldon09, johnston07, mandelbaum08}.  These studies have yielded precise measurements of the clusters' mass-richness relation, which connects the number of galaxies observed to be in the cluster (its richness) to the underlying halo mass.  

Recently, the SZ signal of the MaxBCG clusters has been measured by the Planck satellite \citep{planck_maxbcg}.  The result implies a mass normalization of the clusters that is $\gtrsim 25$\%  lower than that determined using shear.  This difference is significantly larger than the measurement errors quoted in the studies, and represents a puzzle with important implications for both CMB and shear analysis techniques as well as for cluster physics and cosmology.

The discrepancy in the mass-richness relations implied by the weak lensing shear and SZ analyses can be due either to systematic errors in the shear or CMB measurements, or to incorrect  assumptions when comparing the results.   The probability of errors in the basic measurements is small, due to the fact that both the shear and SZ results have been corroborated by additional investigations.

The Planck results have been supported by CMB data from the Wilkinson Microwave Anisotropy Probe (WMAP) satellite  \citep{draper11}.  Furthermore, an analysis of the SZ signal of clusters observed in X-ray data show good agreement between the mass estimates derived from the CMB and the X-ray analyses \citep{planck11}.  It is important to note that while the X-ray and CMB data are entirely independent, they both measure the ICM gas and therefore the interpretations of the signals are subject to the same assumptions about cluster profiles and gas physics.

The separate weak lensing shear analyses of \cite{johnston07} and \cite{mandelbaum08} yield cluster masses that agree at the level of $\sim 5$\%, as discussed in Appendix A of \cite{rozo09} where the measurements are combined into a single relation with appropriate uncertainty, and small systematic differences between the two analyses are taken into account.  However, the two works use similar shear measurement techniques (i.e. weighted second moments of the galaxies' flux distributions, corrected for shape responsivity using the results of \citealt{bernstein02}) and similar background galaxy selections, making the analyses not completely independent.  The lensing measurements are converted to cluster mass using the fact that the tangential shear at a distance $R$ from the cluster center (on the plane of the sky) equals the critical density of the universe at the cluster's redshift, $\Sigma_{crit}$, multiplied by $\Delta \Sigma$: the average projected surface mass density inside a circle of radius $R$ around the cluster, minus the average projected surface mass density at radius $R$.  The shear measurements thereby yield cluster profile shapes, which are fit with Navarro-Frenk-White (NFW) radial profiles \citep{nfw} to determine the cluster masses.  (Only profile shapes, not amplitudes, are probed by weak lensing shear;  magnification measurements do not suffer from this complication, which is known as the mass-sheet degeneracy.)  The fact that the \cite{johnston07} and \cite{mandelbaum08} results are consistent is an important check of the data analysis;  however, because the analysis techniques are similar, the results are subject to similar systematic errors (for example, biases in galaxy shape measurement due to intrinsic properties of the galaxies and the image quality, \cite[see, e.g.,][]{step2, great08_results}).

This work constrains the mass-richness relation of the MaxBCG clusters by measuring the lensing magnification of quasars.  
\cite{agn_magn} demonstrated that the lensing magnification of type I quasars can be measured using the relation between the objects' variability amplitudes and their luminosities.  A measurement of the variability of a quasar yields an estimate of its intrinsic luminosity, which can be compared to its observed luminosity to provide a measurement of its magnification.  

Lensing magnification, like lensing shear, measures the total projected mass of the galaxy cluster acting as a lens.  These two measurement techniques, however, are subject to very different systematic errors.  This work is therefore an important test of the weak lensing shear analyses, checking that the procedures used to measure and interpret shape distortions do not suffer from a large systematic error that could dominate the discrepancy in the mass normalization of the MaxBCG clusters.  Such an independent measurement of the cluster masses is a critical step in understanding and resolving the tension between the gas-based and lensing-based analyses.

\section{Data}

We measure quasar variability using the Palomar-QUEST RG-610 data set, as described in \cite{agn_magn} (hereafter called Paper I).  The variability data are measured and calibrated identically to the data set in that work.  In short, the Palomar-QUEST Variability Survey has observed 30,000 square degrees of sky, imaged multiple times in a broad, red, optical filter (RG-610).  The survey reaches a depth of mag 19.5 in each exposure (under good conditions).  After photometric calibration and strict quality cuts, the data exhibit a systematic error of typically 2\%.

In the current work, the quasar sample is taken from the catalog of quasar properties presented in \cite{shen11}.  The catalog consists of 105,783 quasars from SDSS data release 7 \citep{sdss7}, and includes black hole mass estimates, luminosity measurements at 1350\AA, 3000\AA, and 5100\AA, as well as flags marking, for example, broad absorption line (BAL) quasars and quasars observed in the radio band.  The variability-luminosity relation is seen in radio-quiet quasars, whose variability is thought to relate to the behavior of accretion disk instabilities \citep[e.g.,][]{collier01, vandenberk04, devries05, adam, wilhite08, bauer09a, meusinger11}.  Therefore, we do not use in our analysis quasars that have observed radio emission, which implies the existence of a jet which can be an additional source of optical variability \citep[e.g.,][]{kelly09}.  We also eliminate from the sample those objects flagged as BAL quasars, as the outflows from these objects may affect the observed variability properties.  We investigate the effect of these cuts in section \ref{cut_effects_section}.  We use the catalog's luminosities at 3000\AA\ as the luminosity measure in the variability-luminosity relation, as that wavelength exhibits continuum luminosity from the accretion disk (similar to the luminosity at 2500\AA, which was the measure adopted in Paper I).   We use the mass estimates from the catalog as well, when normalizing the variability data as described below.

\section{Method}

We use the technique presented in Paper I to measure the lensing magnification of type I quasars using the quasar variability-luminosity relation.  We briefly describe the method below.  For further explanation, we refer the reader to Paper I.

\subsection{Measuring the Variability-Luminosity Relation}

The basis of the magnification measurement is the empirical correlation between variability amplitude and luminosity that has been seen in large ensembles of type I quasars (\citealt{vandenberk04}, \citealt{bauer09a}).  This relation allows a measurement of the variability amplitude of a quasar to act as a measurement of the quasar's intrinsic luminosity.  The measured luminosity of the object, divided by this intrinsic luminosity, yields an estimate of the quasar's magnification.  Each magnification measurement has a large error, corresponding to the large scatter in the variability-luminosity relation.  We can average the magnification measurements of many quasars, however, to obtain a statistically significant result.

We measure the variability of quasars using the statistic 
\begin{equation}
V = \sqrt{(\Delta m)^{2} - \sigma^{2}}
\label{v_equation}
\end{equation}
which is similar to the structure function.  $\Delta m$ is the magnitude difference between two measurements of a single quasar, and $\sigma$ is the error on those photometry measurements.  $V$ is measured for each pair of magnitude measurements of a quasar;  $N$ observations of one object will yield $N(N-1)$ measurements of $V$.

Quasar variability amplitude has been seen to depend on a number of factors besides luminosity, for example the time lag between measurements, the quasar's estimated black hole mass, and wavelength of the observations (e.g., \citealt{vandenberk04}, \citealt{devries05}; \citealt{wilhite05}; \citealt{wilhite08}; \citealt{bauer09a}; \citealt{macleod10}; \citealt{meusinger11}).  In order to measure a tight correlation between the variability and luminosity, it is necessary to take into account the variability amplitude's dependence on these other parameters.  This is done by dividing the quasars' $V$ measurements into bins according to black hole mass $M$, wavelength of observation $\lambda$, time lag $\tau$, and luminosity $L$.  Each multi-dimensional bin yields an average $V$ value;  the difference in these average $V$s is due to the dependence of $V$ on mass, wavelength, time lag, and luminosity.  For each choice of ($M$, $\lambda$, $\tau$) bins there exists a $V-L$ relation whose amplitude is related to the dependence of $V$ on $M$, $\lambda$, and $\tau$, but whose slope is due to the relation between $V$ and $L$.  We determine a zero point for each ($M$, $\lambda$, $\tau$) bin, additive in log($V$), such that the different bins' $V$s are normalized to the same amplitude.  The variability's dependence on $M$, $\lambda$, and $\tau$ is therefore removed, and the normalized data demonstrate how $V$ scales with $L$.  
Explicitly, the variability-luminosity relation is of the form
\begin{equation}
\log(V) = C_{0} + \delta C(M, \lambda, \tau) - \alpha \times \log(L_{\mathrm{meas}}).
\label{v_vs_l_equation_a}
\end{equation}
where $C_{0}$ is a constant to be fit using the data, and $\delta C(M, \lambda, \tau)$ 
are the zero points that normalize the different bins' $V$s to the same amplitude.  
We then define the normalized variability $V_{\mathrm{norm}}$ such that 
\begin{equation}
\log(V_{\mathrm{norm}}) = \log(V) - \delta C(M, \lambda, \tau).
\label{vnorm_def}
\end{equation}

Because the Palomar-QUEST data use a single optical filter, the central rest-frame wavelength $\lambda$ at which we observe a quasar at redshift $z$ is simply $\lambda_{\mathrm{RG610}} \times (1+z)$, where $\lambda_{\mathrm{RG610}}$ is the central wavelength of the RG610 filter.  Each measurement includes flux from a range of rest frame wavelengths because the filter is very broad.  However, the observed wavelength range scales uniquely and monotonically with redshift.  Therefore, binning the quasars in redshift $z$ is equivalent to binning them in $\lambda$.  If quasar variability evolves with redshift in addition to depending on observed wavelength, then using single-filter data and binning in either $z$ or $\lambda$ conflates the two effects.  Recent studies measure no significant dependence of variability on redshift \citep{macleod10, meusinger11}, implying that by binning in $z$ we are primarily capturing the change of variability amplitude with rest frame wavelength.

The bin limits in luminosity, black hole mass, redshift (in place of wavelength), and time lag are given in Table \ref{bin_limits_table}.

\begin{table}
\begin{center}
\begin{tabular}{|l|l|l|l|}
\hline
$\tau$ &  $M$ & $z$ & $L$ \\
\hline
1  & 0.5 & 0.80 & 45.00 \\
5  & 4 & 1.15  & 45.25 \\
10 & 8 & 1.40  & 45.50 \\
20 & 12 & 1.65 & 45.75 \\
60  & 20 & 1.90 & 46.0 \\
130  & 30 & 2.20 & 46.25  \\
160 & 75 & & 46.50 \\
220 & & & \\
400 & & & \\
\hline
\end{tabular}
\end{center}
\caption{Bin limits used in determining the normalization constants.  Measurements with values outside the limits are not considered.  Units of time lag $\tau$: 
days; black hole mass $M$: $10^{8} \times M_{\odot}$, luminosity $L$: $\frac{\mathrm{erg}}{\mathrm{s} \cdot \mathrm{Hz}}$.}
\label{bin_limits_table}
\end{table}

\subsection{Data Cuts}

We make similar quality cuts to the data as in Paper I.  The calibrated Palomar-QUEST Survey data include 819,679 measurements of 13,782 objects from the DR7 quasar catalog of \cite{shen11}, after removing BAL quasars and objects with radio detections.  We further wish to eliminate from the sample quasars that show unusual variability behavior, as they are less likely to follow the variability-luminosity relation seen for the ensemble.  For example, a quasar may have a jet which is not bright enough to be noted in the catalog, but still contributes to the optical variability.  Because the variability mechanism for this object would differ from that of the ensemble, we would not expect it to follow the same variability-luminosity relation.  We therefore eliminate objects that exhibit variability that is significantly larger than typically seen in type I quasars.  In particular, for each object we calculate the structure function 
\begin{equation}
SF = \sqrt{<(\Delta m)^{2}> - <\sigma^{2}>}
\label{sf_equation}
\end{equation}
where angular brackets indicate the mean.  $\Delta m$ and $\sigma$ are as in equation \ref{v_equation}.  The structure function $SF$ is calculated twice for each object: once including all measurement pairs $\Delta m$ with time lag less than 10 days in the rest frame, and once including all measurement pairs $\Delta m$ with rest frame time lag greater than 100 days.  If the short timescale $SF$ is larger than 0.1 or the long timescale $SF$ is larger than 0.5, the object is discarded from the sample.  These cut values are motivated by the variability of large ensembles of type I quasars and blazars measured in \cite{blazar_paper}.  We thereby eliminate 1237 objects, leaving 12,545 quasars with 737,730 measurements.  Note that this cut also eliminates objects that may be poorly measured or calibrated in the survey.  %, and that this is an unbiased cut with respect to lensing analyses because the variability amplitude is unchanged for lensed and unlensed objects.

Many objects are cut because they do not lie inside the bin limits given in Table \ref{bin_limits_table}.  For example, there exists a long, low-level tail of measurements with time lags larger than 400 days.  Such tails are difficult to calibrate well, and should be eliminated from the sample.  After binning the data in time lag, redshift, mass, and luminosity we are left with 463,437 measurements of 8499 quasars.

To further eliminate poorly measured data from the sample, we cut any measurement pairs $\Delta m$ that exhibit variability greater than one magnitude, as this is unexpected in type I quasars over time lags studied in this work \citep[see][]{blazar_paper}.  This cuts only 17 measurement pairs.

As in Paper I, we consider only measurement pairs for which we observe significant variability, such that $V > \sigma$ in equation \ref{v_equation}.  This cut eliminates much of the data, but is necessary in order to ensure that the variability measurements are not dominated by noise.  
%We note again that because variability amplitude is unchanged by gravitational lensing, a cut on variability amplitude is unbiased with respect to lensing magnification.  
The resulting data set includes 166,101 measurements from 7713 quasars.

When calculating the zero points for each multi-dimensional bin of mass, wavelength, and time lag, we insist that there exist 30 measurements in each bin such that the zero point is well-determined.  If a bin has fewer members, it is not considered in the analysis.  This criteria leaves 165,845 measurements of 7712 quasars.

Finally, because not all quasars have the same number of measurements, and because a magnification measurement is calculated from each measurement pair (see equation \ref{mu_equation}), a minority of quasars with a large number of measurements may have a disproportionate influence on the results.  We therefore include in the analysis no more than 20 measurement pairs from each quasar.  This cut produces our final data set, with 97,247 measurements of 7712 quasars.

The effects of these cuts on the results are investigated in section \ref{cut_effects_section}.

\subsection{Measuring Quasar Magnification}

After the $V$ measurements are normalized, the variability-luminosity relation is measured by fitting the parameters $\alpha$ and $C$ in the equation
\begin{equation}
\log(V_{\mathrm{norm}}) = C_{0} - \alpha \times \log(L_{\mathrm{meas}}).
\label{v_vs_l_equation}
\end{equation}
where $L_{\mathrm{meas}}$ is the quasar luminosity at 3000\AA, and the relation is fit using the entire quasar sample.  Then, assuming that the mean magnification of the entire quasar sample is unity, the magnification is given by
\begin{equation}
\mu = L_{\mathrm{meas}} \left(\frac{V_{\mathrm{norm}}}{10^{C_{0}}}\right)^{1/\alpha}.
\label{mu_equation}
\end{equation}
Equation \ref{mu_equation} is used to calculate a magnification estimate from 
each $V$ measurement in the data set.   
We can assume that the mean magnification of our quasar sample is indeed unity, as it has been shown in ray-tracing simulations that the mean magnification across the sky is  unity \citep{takahashi11}, and the current data set spans thousands of square degrees of sky and therefore should cover a fair sample of large scale structure.  

The error in a magnification measurement reflects the scatter in the variability-luminosity relation, which is due to measurement uncertainties and also to details of quasar variability that are not well understood.  The error on each measurement is larger than the magnification signal;  therefore, we must average many magnification measurements in order to obtain a significant result.  In order to combine measurements with similar expected magnification, we bin the measurements by their distance (scaled by the cluster radius) from the nearest galaxy cluster in the MaxBCG catalog.  

\subsection{Constraining Galaxy Cluster Profiles}

The MaxBCG catalog contains 13,823 clusters detected in the SDSS, with redshifts between 0.1 and 0.3 and masses greater than roughly $10^{14} M_{\odot}$.  The catalog is estimated to be 90\% pure and 85\% complete in its redshift range and 7500 square degree footprint.  

To measure the average profile of the clusters in the catalog, we average the magnification signal of quasars close to the clusters.  To rescale the clusters' magnification signal so that we can effectively stack clusters with very different richness, we consider the magnification as a function of distance from the line of sight to the cluster, divided by $R_{200}$ of the cluster, where $R_{200}$ is the radius inside which the density is 200 $\times \rho_{\mathrm{crit}}$, the critical density of the universe at the redshift of the cluster.  In order to determine $R_{200}$, and to calculate the magnification expected from the clusters, we must assume a cluster mass model.  We assume that each cluster has a NFW profile \citep{nfw}:
\begin{equation}
\rho(r) = \frac{\delta_{c} \rho_{\mathrm{crit}}}{(r/r_{s})[1+(r/r_{s})]^{2}}
\label{nfw_equ}
\end{equation}
where $c$ is the concentration parameter of the profile, $\delta_{c} = \frac{200}{3} \frac{c^{3}}{\mathrm{ln}(1+c) - c/(1+c)}$, and $r_{s} = R_{200}/c$.  The total mass inside $R_{200}$ is $M_{200}$.  

In order to specify a unique model for the cluster masses, we must adopt a mass-concentration relation and a mass-richness relation.  Given these assumptions, the richness parameter of each cluster in the MaxBCG catalog fully determines the mass profile model.  We assume the mass-concentration relation from \cite{duffy08}:
\begin{equation}
c(M_{200},z) = 5.71 \left(\frac{M_{200}}{2 \times 10^{12} h^{-1} M_{\odot}}\right)^{-0.084} ( 1 + z )^{-0.47} .
\end{equation} 

The NFW magnification profile can be calculated analytically using the profile parameters, as explained in \cite{schneider04}.  

The mass richness relation is the quantity we wish to constrain.  We adopt the form of the relation used in \cite{rozo09}:
\begin{equation}
M_{200}(N_{200}) = \frac{1}{1.022} e^{A} \left(\frac{N_{200}}{20} \right)^{B} 10^{14} h^{-1} M_{\odot}
\label{rozo_equ}
\end{equation}
where $N_{200}$ is the number of red sequence galaxies observed to be in the cluster that are above a limiting brightness and within a certain radius, and is provided in the MaxBCG catalog (see \citealt{maxbcg} for details).  

Our fitting procedure involves choosing $A$ and $B$, calculating the expected magnification of the quasar sample in logarithmic bins versus radius, and comparing it to the data in order to calculate the probability that the data are consistent with the model.  We renormalize the expected magnification values such that the mean over the entire quasar sample is equal to one;  this typically shifts the values by only about 0.4\%.  We choose $A$ and $B$ from a grid of points with spacing 0.01 in each dimension, from ($A,B$) of (0.1,0.8) to (0.8,1.3).   

\section{Results}

\begin{figure}
\begin{center}
\plotone{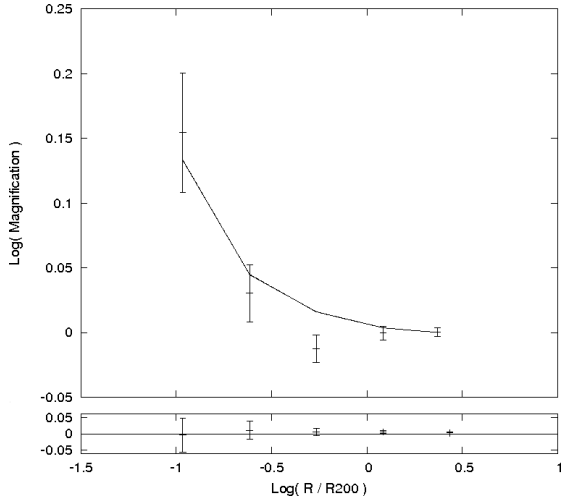}
\end{center}
\caption{Magnification versus transverse distance to the nearest cluster (scaled as $R/R_{200}$).  Stacked signal using all MaxBCG clusters.  The solid line shows the expected magnification profile, calculated assuming a mass-richness relation of the form given in equation \ref{rozo_equ}, with $A=0.49$ and $B=0.95$, which has a 16\% probability of being consistent with the data.  The bottom panel shows a profile measured after randomizing the measurements, as described in the text, which is consistent with zero.}
\label{profile_figure}
\end{figure}

An example plot of the expected and measured magnification, versus scaled distance from the nearest cluster, is shown in the main panel of Figure \ref{profile_figure}.  The bins' expected magnifications are connected by lines for clarity.  The values of $A$ and $B$ are 0.49 and 0.95, respectively.  To avoid the fit quality from being dominated by properties outside of the main cluster halo we consider five bins, out to radius $3 \times R_{200}$.  This particular combination of $A$ and $B$ provides the best fit to the data, with a chi square of 8.0 with 5 degrees of freedom, corresponding to 16\% probability of being consistent with the data.  

The sources and detailed properties of the scatter in the variability-luminosity relation, which is the main source of error in the magnification measurements, are not well understood.  Therefore, we calculate the magnification errors empirically using the distribution of measurements.  We calculate the error for each radial bin using bootstrap resampling, with 500 subsamples per bin.  Specifically, for each bin we randomly choose N objects from the data in that bin, where N is the number of data points in the bin.  Because we choose randomly with replacement, the bootstrap sample is different from the original data sample in that some values are missing and others are repeated.  We construct 500 such bootstrap samples, and take the error to be the range that encloses 68\% of the samples.  We have tested that the errors are insensitive to the exact number of bootstrap subsamples, and that randomly shuffling the magnification measurements (i.e. replacing each measurement with another from the data set in a random manner) produces a profile consistent with zero to within $1\sigma$.  An example instance of the profile using randomized measurements is shown in the bottom panel of Figure \ref{profile_figure}.

\subsection{Mass-Richness Relation \label{m_n_relation_section}}

\begin{figure}
\begin{center}
\plotone{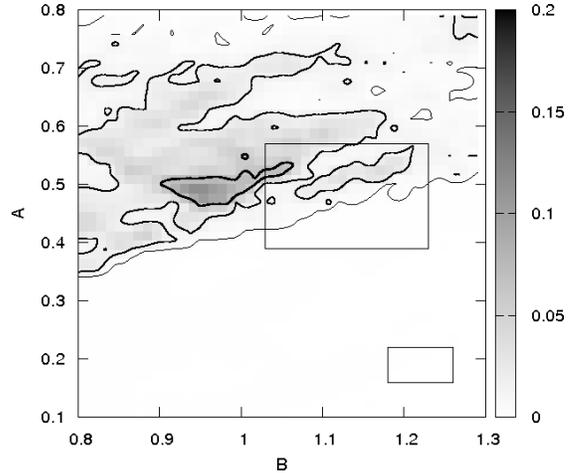}
\end{center}
\caption{Mass-richness relation parameters A versus B, with gray scale corresponding to the probability of agreement with the data.  The contours mark 68.3\%, 95.5\%, and 99.7\% confidence regions around the best-fit values.  Boxes represent (from top to bottom) the $1\sigma$ regions of \cite{rozo09} and \cite{planck_maxbcg}.}
\label{3dplot}
\end{figure}

Figure \ref{3dplot} shows the mass-richness parameters $A$ versus $B$, with the gray scale corresponding to the probability that the mass model is consistent with the data.  The contours mark the 68.3\%, 95.5\%, and 99.7\% (in order of decreasing line thickness) confidence regions around the best-fit parameters.  The upper and lower boxes delineate the $1\sigma$ results of \cite{rozo09} and \cite{planck_maxbcg}, respectively.  The best-fit model parameters, with marginalized 68\% statistical errors, are $(A, B) = (0.49^{+0.01}_{-0.01}, 0.95^{+0.03}_{-0.01})$.  The best-fit values have a 16\% probability of agreeing with the data, and yield the profile shown in Figure \ref{profile_figure}.

\cite{rozo09} combine previous shear analyses to conclude ($A,B$) = ($0.48 \pm 0.07 (\mathrm{stat}) \pm 0.06 (\mathrm{sys}), 1.13 \pm 0.09 (\mathrm{stat}) \pm 0.05 (\mathrm{sys})$).  These systematic errors include the effects of cluster miscentering on the average shape of the haloes, due to incorrect identification of the central galaxies as discussed in detail in \cite{johnston07}.  The Planck SZ measurements are translated to ($A,B$) values using the fit results from Table 2 of \cite{planck_maxbcg} (using the \citealt{rozo09} mass calibration) for the $\tilde{Y}_{500}-N_{200}$ relation, where $\tilde{Y}_{500}$ is the SZ signal:  the integral of the electron pressure over the volume of the cluster, times a geometric factor, scaled to a redshift of zero so that the measurements of all of the clusters are directly comparable.  The conversion between $\tilde{Y}_{500}$ and $M_{500}$, as put forth in \cite{arnaud10} and adopted in \cite{planck_maxbcg} is such that $\tilde{Y}_{500} \propto M_{500}^{5/3}$.  $M_{500}$ is the mass of the cluster inside a radius within which the average matter density is 500 times the critical density of the universe at the redshift of the cluster.  Making the simplifying assumption that the conversion from $M_{500}$ to $M_{200}$ does not introduce error into the mass-richness relation parameters, 
the fractional error on the $M_{500}-N_{200}$ parameters from the SZ analysis will be the same as the fractional error on the equivalent $M_{200}-N_{200}$ relation parameters.  We can therefore propagate the errors from the $\tilde{Y}_{500}-N_{200}$ relation to approximate errors in an $M_{200}-N_{200}$ relation.  \cite{planck_maxbcg} state that their results are in agreement with the mass-richness relation given in \cite{johnston07}, except with an amplitude smaller  by 25\%. We therefore use the \cite{johnston07} results to determine the Planck-based value of $A = 0.19 \pm 0.03$.  The Planck-based $B$, calculated from their parameter $\alpha$, is 1.22 $\pm$ 0.04.

\subsection{Effects of Data Cuts \label{cut_effects_section}}

To test that the cuts made to the data set do not bias the cluster mass measurements, we repeat the analysis using data samples with modified selection criteria.  

In the main analysis, we eliminate from our sample quasars detected in the radio band by the FIRST Survey \citep{becker95}, which are flagged in the value-added quasar catalog.  Because jets produce radio emission and also affect quasar optical variability, we expect those quasars with radio emission to introduce extra scatter into the variability-luminosity relation, creating noise in the cluster mass measurements.  Because there are bound to be quasars in the sample with radio fluxes just below the FIRST detection limit, it is important to ensure that radio emission is not associated with strong biases in the mass measurements.  To explicitly test the effects of radio-detected quasars on the mass measurements, we repeat the above analysis while including quasars with radio fluxes measured to be less than 10 mJy at 6 cm (as given in the \cite{shen11} quasar catalog).  This comprises 52\% of the radio-detected quasars in the catalog.  We find that similar results when including the radio detections.  Specifically, with 68\% confidence, $(A, B) = (0.50^{+0.05}_{-0.01}, 0.98^{+0.04}_{-0.06})$.  The inclusion of the radio-detected quasars enlarges the confidence region to higher mass normalizations, as expected by the fact that radio-loud AGN show higher variability than those that are radio-quiet.  The effect, however, does not significantly bias the results.

Including only variability measurements $V$ that are larger than $2\sigma$ rather than $1\sigma$, eliminating highly variable objects with long- and short-term $SF$ 25\% larger or smaller than the standard cut, and allowing only 10 measurements per objects rather than 20, each changes the best-fit $(A,B)$ by less than the 68\% errors.  Eliminating the typically least-variable measurements, by discarding the first two bins in time lag, similarly has no significant effect on the results.

Including the 5258 AGN marked as BAL quasars in the value-added catalog decreases the best-fit $A$ and $B$ along with the probability of the best-fit model representing the data.  The best-fit region lies along the same degeneracy direction as is visible in Figures \ref{3dplot} and \ref{3dplot_ellip}, although the highest probability values are at the low-$B$ range of the parameter space.  Because including known BAL quasars in the sample significantly lowers the best-fit $(A,B)$ values, unidentified BAL quasars remaining in the sample may be artificially lowering the best-fit mass-richness parameters for our main quasar sample.  Correcting for such contamination would serve to improve the agreement with the \cite{rozo09} mass-richness relation.

To examine the sensitivity of our results to the chosen concentration relation, we repeat the analysis using an alternate relation, from \cite{mandelbaum08}:
\begin{equation}
c(M_{200},z) = \frac{4.6}{1+z} \left(\frac{M_{200}}{1.56 \times 10^{14} h^{-1} M_{\odot}}\right)^{-0.13}.
\end{equation} 
This relation changes the results by less than the 68\% errors.

\subsection{Effects of Cluster Miscentering}

The MaxBCG catalog is known to suffer from miscentering uncertainties.  The cluster centers are given as the location of the brightest cluster galaxy (BCG).  In some cases, the BCG may be offset from the true halo center.  More often, the BCG is misidentified.  Such errors in the determination of the halo center will lead to errors in the stacked halo profile, in particular making the stacked profile shallower than expected.  This effect can show up as an artificial change in concentration depending on the innermost radius probed \citep{mandelbaum08}.  \cite{johnston07} studied this problem in detail using simulations, resulting in a mass-dependent probability that a cluster is offset from its true center by a given amount.

The fact that our results are robust to changes in the cluster concentration relation implies that they may not be strongly affected by miscentering.  To investigate this further, we implement the miscentering prescription of \cite{johnston07}, offsetting the centers of a random minority of the clusters.  This procedure likely overestimates the true miscentering properties of the catalog \citep{mandelbaum08}.  It is not meant to correct for the cluster center misidentification, since the randomly-chosen offsets will almost always bring the considered coordinates farther away from the true cluster center.  Instead, applying this prescription is meant to show the effects of such center shifts on the results.  The procedure adds noise to the results, while moving the best-fit region along the direction of $(A,B)$ degeneracy towards lower best-fit values.  The precise best-fit region is sensitive to the random choice of cluster offsets;  different realizations of the procedure yield different results.  Because the best-fit values consistently lie along the line of $(A,B)$ degeneracy, miscentering is unable to bring our results in line with the predictions of \cite{planck_maxbcg}.  If it were possible to correct for cluster miscentering, our best-fit $A$ and $B$ would both increase.  Such behavior would not ease the tension between the weak lensing and SZ results.

\subsection{Effects of Cluster Orientation}

Brightest cluster galaxies (BCGs) have been observed to have major axes that are typically aligned with that of their host dark matter haloes \citep{hashimoto08, niederste10}.  The morphologies of the clusters, as opposed to their axes of orientation, are not seen to be related to the morphologies or orientations of the BCGs.  Because of this connection, clusters with BCGs that appear elliptical on the plane of the sky are more likely to be oriented face-on than those with BCGs that appear spherical (because an apparently-spherical BCG may be an elliptical galaxy with major axis close to the line of sight). 

\cite{marrone11}, in comparing the weak lensing and SZ signals from 18 galaxy clusters, measured a scatter around the mean relationship between shear-based mass and SZ signal that was dependent on cluster morphology and orientation.  The clusters' morphology was determined using X-ray data, and undisturbed clusters yielded systematically larger weak lensing masses than disturbed clusters with similar SZ signal.   The clusters' orientation was approximated by the ellipticity of their BCGs;  those with axis ratio larger than 0.85 (i.e., clusters preferentially oriented along the line of sight) also had systematically larger weak lensing masses for a given SZ signal.  These results agree with the expected relationship between halo orientation and bias in cluster weak lensing mass measurements \citep{corless07}.

\begin{figure}
\begin{center}
\plotone{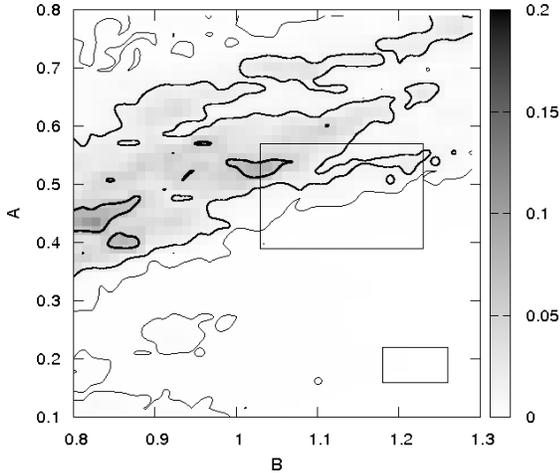}
\end{center}
\caption{Mass-richness relation parameters A versus B, with gray scale corresponding to the probability of agreement with the data, for the ``face-on" cluster sample.  The contours mark 68.3\%, 95.5\%, and 99.7\% confidence regions around the best-fit values.  Boxes represent (from top to bottom) the $1\sigma$ regions of \cite{rozo09} and \cite{planck_maxbcg}.}
\label{3dplot_ellip}
\end{figure}

Because X-ray measurements are not available for the majority of the MaxBCG catalog, it is difficult to estimate the morphology of our cluster sample.  The ellipticities of the BCGs are available from the SDSS CasJobs server\footnote[1]{http://cas.sdss.org/astrodr7/};  we use these as a proxy for the orientation of the MaxBCG clusters.  We repeat our analysis using only those MaxBCG clusters for which the BCG has axis ratio less than 0.85.  By matching the MaxBCG cluster central position with the SDSS DR7 galaxy table, with a spatial matching tolerance of 1", we find 10,248 BCG galaxies with axis ratio, calculated by fitting an elliptical deVaucouleurs profile to the r band SDSS data, of less than 0.85.  
The mass-richness parameters determined for this ``face-on" sample is shown in Figure \ref{3dplot_ellip}.  The results agree within $1\sigma$ with those of the whole sample, although the best-fit region extends to significantly lower $A$ and $B$ values.

\begin{figure}
\begin{center}
\plotone{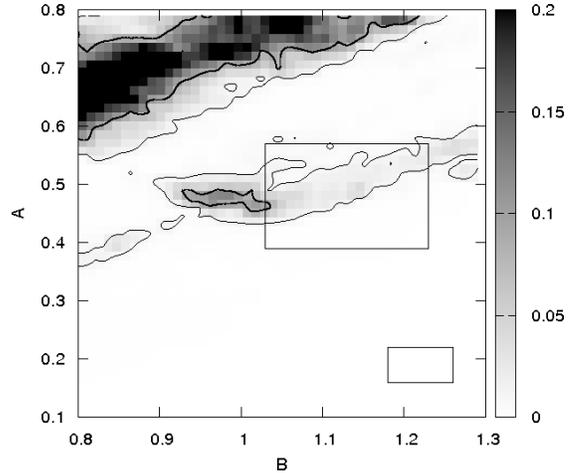}
\end{center}
\caption{Mass-richness relation parameters A versus B, with gray scale corresponding to the probability of agreement with the data, for the ``line-of-sight" cluster sample.  The contours mark 68.3\%, 95.5\%, and 99.7\% confidence regions around the best-fit values.  Boxes represent (from top to bottom) the $1\sigma$ regions of \cite{rozo09} and \cite{planck_maxbcg}.}
\label{3dplot_round}
\end{figure}

To compare with the face-on cluster sample, we construct a MaxBCG subsample including the 10,248 clusters with the roundest BCGs.  There is substantial overlap between this subsample and the ``face-on'' one used above;  this subsample includes clusters with BCG axis ratio greater than $\sim0.68$.  Nevertheless, this ``line-of-sight'' subsample can serve as a comparison to highlight the effects of cluster orientation on the measurement.  
The mass-richness fit results for the ``line-of-sight" subsample are shown in  Figure \ref{3dplot_round}.  The best-fit relation indicates significantly higher mass normalization for these ``line-of-sight" clusters than for the whole sample and the ``face-on" subset.  Furthermore, the probability that the model represents the data is much higher for the best-fit parameters in the ``line-of-sight" subsample (64\%) than in the ``face-on" subsample (14\%), implying that correlated structure in the plane of the sky in face-on clusters decreases the quality of a spherical NFW fit.  We note that the subsamples' best-fit regions extend beyond our explored parameter space, although the probed region is sufficiently large for the purpose of comparing the subsets' general behavior.

\begin{table}
\begin{center}
\begin{tabular}{|l|l|l|l|l|l|}
\hline
 & $A_{\mathrm{best}}$ & $B_{\mathrm{best}}$ & $\mathrm{P_{best}}$ & $\mathrm{P_{Rozo}}$ & $\mathrm{P_{Planck}}$ \\
\hline
Total Sample & $0.49^{+0.01}_{-0.01}$ & $0.95^{+0.03}_{-0.01} $& 0.16 & 0.11 & 7.0e-06\\
Face-on & $0.42^{+0.10}_{-0.01+}$ & $0.81^{+0.23}_{-0.01+} $ & 0.14 & 0.08 & 1.5e-05\\
Line-of-Sight & $0.69^{+0.01}_{-0.03+}$ & $0.83^{+0.07}_{-0.03+} $ & 0.64 & 0.04 & 5.7e-08\\
\hline
\end{tabular}
\end{center}
\caption{Best-fit mass-richness parameters $A$ and $B$ and the probability that the magnification model with these parameters represents the data.  $\mathrm{P_{Rozo}}$ and $\mathrm{P_{Planck}}$ are the total probability within the $1 \sigma$ error regions of the results from \cite{rozo09} and \cite{planck_maxbcg}, respectively, normalized to the probability in the region from $(A,B)$ (0.1,0.8) to (0.8,1.3). The ``face-on" and ``line-of-sight" subsamples' error regions extend outside of our parameter space.}
\label{results_table}
\end{table}

\section{Discussion and Conclusions}

We constrain the mass-richness relation of the MaxBCG galaxy cluster catalog by measuring the lensing magnification of background quasars.  The magnification is determined by measuring the quasars' variability and using the correlation between their variability and luminosity to constrain the objects' intrinsic luminosity.  Comparing this intrinsic luminosity to the measured value yields a measurement of the quasar's magnification.  Because each magnification measurement has low signal to noise, we stack the measurements to generate a single radial magnification profile as a function of distance from a cluster.

Each galaxy cluster is assumed to have a NFW density profile, with a mass that is related to its richness $N_{200}$ that is given in the MaxBCG catalog.  The mass-richness relation is assumed to be of the form given in equation \ref{rozo_equ}.  Assuming values for the relation's normalization $A$ and slope $B$, the magnification expected for the quasars, due to the galaxy clusters, can be calculated.  The probability of the assumed mass-richness parameters is taken to be the agreement between the measured magnification and the expected values calculated using the clusters' mass-richness relation and the lensing properties of NFW haloes.  These probabilities are used to determine the level of agreement between the data and the model, and between the data and the results of \cite{rozo09} and \cite{planck_maxbcg}.  The results are summarized in Table \ref{results_table}.

The best-fit and 68\% statistical errors for the mass-richness parameters are $(A,B) = (0.49^{+0.01}_{-0.01}, 0.95^{+0.03}_{-0.01})$.  Assuming that the model used to calculate the cluster magnification is accurate (i.e. that the clusters follow NFW profiles which obey a mass-richness relation parameterized by $A$ and $B$), we find 11\% agreement between the magnification results and those of \cite{rozo09}.  In contrast, we find 7e-04\% agreement between the magnification data and the mass-richness relation implied by the SZ results of \cite{planck_maxbcg}.  The magnification results thereby corroborate the weak lensing shear measurement of the MaxBCG mass-richness relation, and strongly disfavor the lower mass normalization implied by SZ and X-ray analyses.  The data, however, are not perfectly represented by the model;  the unnormalized best-fit probability that the data are represented by the model is only 16\% for the best-fit parameters.  We have checked the response of the results to cuts in the data, and to cluster miscentering.  Systematic errors such as cluster miscentering and inclusion of heterogeneous AGN in the sample (such as BAL quasars) serve to decrease $A$ and $B$ simultaneously, along a line of degeneracy visible in Figures \ref{3dplot} and \ref{3dplot_ellip}.  These effects are larger than the statistical errors;  however, correcting for them would not bring the results closer to those of the SZ and X-ray analyses. 

The normalization of the mass-richness relation depends on the orientation of the galaxy clusters with respect to the line of sight.  We split the MaxBCG catalog into two overlapping sets, each of size $\sim75\%$ of the whole catalog, based on the measured ellipticity of their BCGs which can be used as a diagnostic for the cluster halo orientation.  The ``face-on'' subset (with BCG axis ratio less than 0.85) shows behavior similar to the whole sample, with consistent best-fit values although the 68\% confidence region extends along a line of degeneracy to smaller $A$ and $B$ values.  The ``line-of-sight'' subset (with BCG axis ratio greater than 0.68), which has identical statistics (and shares $\sim$ 50\% of its objects with the ``face-on" subset) has a best-fit mass normalization $A$ that is 40\% higher than that of the whole sample.  This is in line with predictions from numerical simulations for the lensing mass bias due to assuming spherical NFW profiles for prolate, line-of-sight oriented clusters \citep{feroz11}.  Although the minority of clusters oriented close to the line of sight fit to a high mass normalization, they are not responsible for the disagreement between the main result and the SZ best-fit region, since both subsamples are incompatible with the results of \cite{planck_maxbcg}.  

The magnification-based measurement of the MaxBCG galaxy cluster mass-richness relation presented in this work is based on an analysis technique with very different systematic errors from those of weak lensing shear.  Because the mass-richness relation of the MaxBCG cluster catalog is consistent when measured using lensing shear and lensing magnification, the discrepancy between the lensing-based cluster masses and the SZ and X-ray ICM-based masses is unlikely to be related to a systematic problem with the shear measurement techniques.  Instead, the cause is likely a physical assumption about the galaxy clusters, made by one or both analyses.  One assumption made by all of the analyses is that the clusters are well-represented by a single, spherical profile shape, while in fact the clusters are triaxial.  Weak lensing shear measurements are expected to have biases due to cluster orientation \cite[see][and references therein]{becker11}, moreover due to the combination of orientation and halo shape (i.e. whether it is prolate or oblate) \citep{corless08, feroz11}.  Our results indicate that although the orientation of the triaxial haloes does affect the mass measurement, it does not significantly bias the average mass-richness relation of the MaxBCG catalog.  Further investigation is needed into assumptions in the lensing and SZ analyses in order to resolve the significant discrepancy between the two methods' measurements of the MaxBCG cluster catalog.

\section*{Acknowledgments}
We thank Eduardo Rozo for suggesting a comparison with the Planck MaxBCG results.  We appreciate the work of 
Peter Nugent and Janet Jacobsen in developing and running the DeepSky pipeline.  
We acknowledge the support of the Spanish Science Ministry
AYA2009-13936, Consolider-Ingenio CSD2007-00060, project
2009SGR1398 from Generalitat de Catalunya, and 
the Marie Curie European Reintegration Grant PERG07-GA-2010-268290.  
The National Energy Research Scientific Computing Center,
which is supported by the Office of Science of the U.S.
Department of Energy under Contract No. DE-AC02-05CH11231,
has provided resources for this project by supporting staff,
providing computational resources and data storage.  
We thank the Office of Science of the Department of Energy
(grant DE-FG02-92ER40704) and the National Science Foundation
(grants AST-0407297, AST-0407448, AST-0407297) for support.

\bibliography{apj-jour,bibliography}

\end{document}